\documentstyle[aps,preprint]{revtex}
\begin{document}
\def \sm {{small-world }}
\def \Sm {{small-world }}
\def \fe {{far edge }}
\def \fes {{far edges }}
\def \N {{\cal N}}
\def \F {{\cal F}}

\title{Characterization and control of small-world networks}
\author{S. A. Pandit\footnote{e-mail: sagar@prl.ernet.in} and
R. E. Amritkar\footnote{e-mail: amritkar@prl.ernet.in}} 
\address{Physical Research Laboratory, Navarangpura, Ahmedabad, 380 
009, India}
\maketitle
\begin{abstract}
Recently Watts and Strogatz have given an interesting model of
small-world networks. Here we concretise the concept of a ``far away''
connection in a network by defining a {\it far edge}. Our definition
is algorithmic and independent of underlying topology of the
network. We show that it is possible to control spread of an epidemic
by using the knowledge of far edges. We also suggest a model for
better advertisement using the far edges. Our findings indicate that
the number of far edges can be a good intrinsic parameter to
characterize small-world phenomena.  
\end{abstract}


%\begin{multicols}{2}

The properties of very large networks are mainly determined by the
way the connections between the vertices are made. At one extreme are
the regular networks where only the ``local'' vertices are
inter-connected and the ``far away'' vertices are not connected while
at the other extreme are the random networks where the vertices are
connected at random.  The regular networks display a high degree of
local {\it clustering} and the average distance between vertices is
quite large. On the other hand, the random networks show negligible
local {\it clustering} and the average distance between vertices is
quite small. The {\it small-world} networks~\cite{Psycho,Play} have
intermediate connectivity properties but exhibit a high degree of
clustering as in regular networks and small average distance between
vertices as in random networks. A very interesting model for
small-world networks was recently proposed by Watts and 
Strogatz~\cite{Watts-Strogatz}. They found that a regular network
acquires the properties of a small-world network with only a very small 
fraction of connections or edges (about $1\%$) rewired to ``far away''
vertices. They demonstrated that several diverse phenomena like neural
networks~\cite{Hopfield}, power grids and collaboration graphs of film 
actors~\cite{Collins-Chow} can be modeled using small-world
networks. Also the spread of 
an epidemic is much faster in small-world networks than in the regular
networks and almost close to that of random networks. 

In this paper we suggest a possible way of characterizing small-world
networks. The basic ingredients of small-world networks are the
``far away'' connections. We introduce a notion of {\it far edges} in a 
network to identify these ``far away'' connections. Our definition of
a {\it far edge} is independent of any underlying topology for a
network and depends only on the way connections or edges are made. We
claim that the rapid spread of an epidemic in small-world network as
found by Watts and Strogatz~\cite{Watts-Strogatz} is due to these far
edges. This allows us to propose a mechanism to control the epidemic
using the same far edges which are responsible for the rapid spread.
We further demonstrate the utility of our notion of far edges by giving 
an better method of advertisement.

Consider a graph (network) with $n$ vertices and $E$ edges. Let
$\N^\nu_{ij}$ denote the number of distinct paths of length $\nu$
between the vertices $i$ and $j$. For a simple graph, $\N^1_{ij}$ is
one if there is an edge between vertices $i$ and $j$ else it is zero.
We now concretise the idea of ``far away'' connections by defining a
far edge. Let an edge $e_{ij}$ between vertices $i$ and $j$ be a far
edge of order $\mu$  if it is an edge for which $\N^{\mu+1}_{ij} = 0$ 
and $\N^l_{ij} \not= 0$ for all $l \leq \mu$.

Fig. 1 shows an example of a far edge of order one. We note that none
of the edges in a completely connected graph are far edges, while all
edges in a tree are far edges of order one. Hence forth we will assume
that a far edge has order one unless stated otherwise.

To generate small-world networks and also other type of networks we
follow the procedure given in Ref.~\cite{Watts-Strogatz}. We start
with a regular network consisting of a ring of $n$ vertices with edges
connecting each vertex to its $k$ nearest neighbours.  Each edge is
rewired with probability $p$ avoiding multiple edges. The $p=1$ case
corresponds to a random network. The networks obtained with $p \approx 
0.01$ correspond to small-world networks~\cite{Watts-Strogatz}.

We have generated several networks from regular ($p=0$) to random
($p=1$) case. For each network we calculate the average path length
$L(p)$ and clustering coefficient $C(p)$. The quantity $L(p)$ denotes the
average length of the shortest path between two vertices, and $C(p)$
denotes the average of $C_v$ over all the vertices $v$, where $C_v$
 is the number of edges connecting the neighbours of $v$ normalized
with respect to the maximum number of possible edges between these
neighbours~\cite{Watts-Strogatz}. Next we determine the far edges in these
networks. Let $\F$ denote the ratio of number of far edges with the
total number of edges. We find that initially, to a good approximation, $\F$ is equal to
$p$ for $p \leq 0.1$ and then it increases slowly till it saturates to a
value of about $0.2$ for $p=1$. It turns out that the number far edges of
order higher than one are negligible.  

In Fig. 2 we plot $C({\cal F})/C(0)$ and
$L({\cal F})/L(0)$ as functions of $\F$. This figure is similar in nature to
the plot of $C(p)/C(0)$ and $L(p)/L(0)$ as functions of $p$ (Fig. 2 of
Ref.~\cite{Watts-Strogatz}). The small-world networks can be
identified as those with $C(p)/C(0) \approx 1$ and $L(p)/L(0) \approx
L(1)/L(0)$. From Fig. 2 we see that  this corresponds to $\F \approx
0.01$. Thus $\F$ can be used as a parameter to characterize networks
which interpolate between regular and random cases. We note that $\F$
is an intrinsic quantity and does not depend on the procedure of 
generating networks and hence should prove to be a better parameter
than $p$. 

To further investigate the importance of far edges, we consider the
problem of spread of an epidemic. Consider an epidemic starting from a
random vertex (seed). We assume that at each time step all the
neighbours of infected vertices are affected with probability one,
which is the most infectious case, and the vertices which are already
affected die and play no further role in the spread of the
epidemic. Here, neighbours of a given vertex means all the vertices
which are joined to it by edges. As found by Watts and
Strogatz~\cite{Watts-Strogatz}, the 
spread of an epidemic in small-world networks is almost as fast as that
in the random case. We propose that the mechanism for the rapid spread 
of epidemic in small-world networks is due to the traversal of the
disease along the far edges. Each such traversal opens a virgin
area for the spread of epidemic leading to a rapid growth. 

Clearly if the far edges are responsible for the rapid growth of
epidemic then we should be able to effectively control the spread by preventing
the traversal of epidemic along the far edges. To test this
hypothesis, we propose the following mechanism to control an
epidemic. We assume that we have sufficient knowledge of the network
and we have identified all the far edges. We note that identification
of far edges requires only the knowledge of vertices and edges and
hence should be possible in many practical situations. Let $\tau$
denote the time steps elapsed between the beginning of the epidemic and its 
detection. Let $m$ denote the number of vertices that can be immunized 
at each time step. To block a far edge we first immunize one of the
two vertices connected by this far edge. Immunization is carried out
by first blocking all the far edges and then immunizing at random. If
the number of far edges is greater than $m$ then blocking all the
far edges will take more than one time step.

In Fig. 3 we show the fraction of vertices affected as a function of
time steps for a small-world network. Curve (a) shows the uncontrolled
spread of the epidemic. Curves (d) and (g) show the spread of epidemic with the
control method suggested above for $\tau = 7$ and $2$
respectively. For comparison we show, by curves (c) and (f), the
epidemic with only random immunization for $\tau = 7$ and $2$
respectively. It is obvious that the far edge control mechanism proposed here
is very effective. For larger $\tau$ some of the far edges are
already traversed by the epidemic, decreasing the efficiency of our
control mechanism. Comparing the far edge immunization and the random
immunization, we find that the far edge immunization decreases the
rate of spread of epidemic more effectively but takes longer time for
completely stopping the spread (See Fig. 3, curves (d) and
(g)). Further, to test the effectiveness of our method we compare the
results with another method of immunization. We order the vertices by
their degree. Immunization is carried out by starting with the vertex
with the largest degree and then going down the degree. The results
for $\tau = 7$ and $2$ are shown as curves (b) and (e) in Fig. 3
respectively. We note that results for immunization using degree are
similar to that of the random immunization.

Let $d$ denote the asymptotic difference between the number of
affected vertices in random and far edge immunization. We plot $d$ as
a function of $m$ for three different values of $\F$ (or $p$) in
Fig. 4. The plot shows that the far edge 
immunization is most effective when $m$ is about half the number of far 
edges. The reason for the decrease of $d$ for large $m$ is that  the
probability that random immunization
blocks a far edge, keeps on increasing as $m$ increases, thereby
decreasing the difference between the two methods. The plot of $d$ as a
function of $\F$ for different values of $m$ is shown in Fig. 5. The
figure shows that the far edge immunization is more effective for
small-world networks. Also from Figs. 4 and 5  it is clear that the far
edge immunization gives a substantial benefit in terms of number of
unaffected vertices in the small-world case and this number can be as large 
as $410$ which is more than $40\%$ of the total number of vertices.  

Now, we consider an interesting model of advertisement. Let $r$ be the
number of vertices or centers from where a product is
advertised. The information about the product spreads by word of mouth 
to the neighbours with the probability $q_t$ where $t$ is the time
elapsed from the initial advertisement. We compare the results of
two different ways of choosing the initial centers. In one way the
centers are chosen at random and in the other they are chosen as
one of the vertex in a far edge. Fig. 6 shows the number of people
informed about the product as a function of $t$. It is clear that the
choice of centers using far edges has definite advantage over that of
random choice. 

To conclude we have introduced the concept of {\it far edges} in
networks. Our definition of a far edge is in accordance with the
intuitive idea of a ``far away'' connection between two vertices. The
advantage of our definition of far edge is that it is 
independent of the underlying topology of the network. Also the
definition is algorithmic in nature, and allows the determination of far 
edges only from the knowledge of vertices and edges. We have also applied
the idea of far edges to the networks which are not generated by the
algorithm given in Ref.~\cite{Watts-Strogatz} and arrived at similar
conclusions~\cite{SAP-REA}.

We have demonstrated the use of far edges in the control of the spread of
an epidemic and the advertisement of products. Our simulations show  that 
the far edges are indeed important in the spread of epidemic,
particularly in the small-world networks. We have shown that the knowledge
of far edges can be fruitfully utilized to control the spread of epidemic
and better advertisement. Our results strongly indicate that the
far edges are the key elements responsible for the special properties
of small world phenomena.

\newpage
\noindent
{\bf Figure Captions:}
\begin{description}
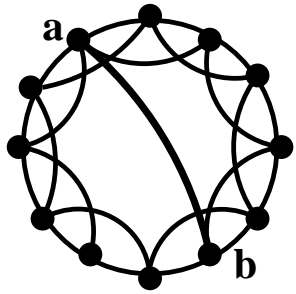
\item [Figure 1] An example of a network consisting of a far edge. The edge 
between vertices `a' and `b' is a far edge of order one.
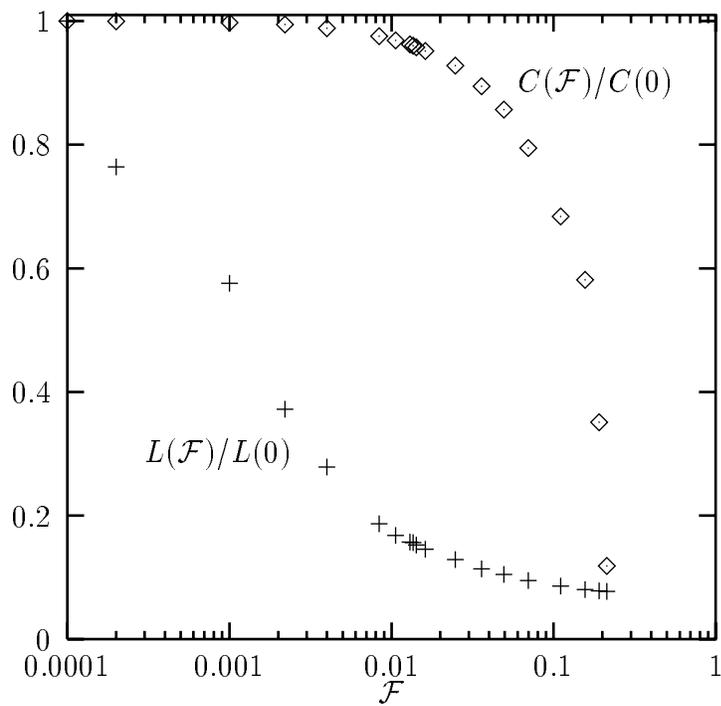
\item [Figure 2] The graph of $C({\cal F})/C(0)$ and $L({\cal
F})/L(0)$ as a function of ${\cal F}$, where $C$ is the clustering
coefficient, $L$ is the average path length and ${\cal F}$ is the
ratio of the number of far edges with the total number of edges. This
figure is similar in nature to the plot of $C(p)/C(0)$ and $L(p)/L(0)$
as functions of $p$. The small-world networks lie around ${\cal F} = 0.01$.
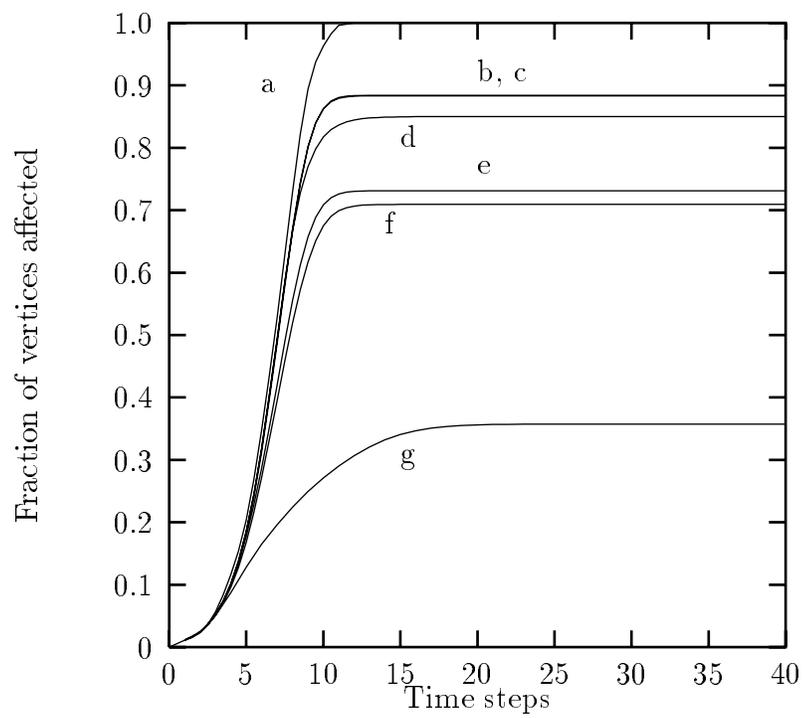
\item [Figure 3] The graphs of fraction of vertices affected as a
function of time steps. The curve (a) 
is the epidemic spread without immunization, the curves (c) and (f)
represent the spread when the random immunization is applied (see text) for 
$\tau = 7$ and $2$ respectively, the curve (b) and (e) shows the spread if the
immunization is carried out for the vertices with highest degree first 
and then in descending degree for $\tau = 7$ and $2$ respectively and
the curves (d) and (g) are the spread when the far edge immunization is
used $\tau = 7$ and $2$ respectively. The simulations are carried out
on a small-world network of 1000  vertices and 10000 edges. The
plotted results are averaged quantities over 500 seeds for epidemic. 
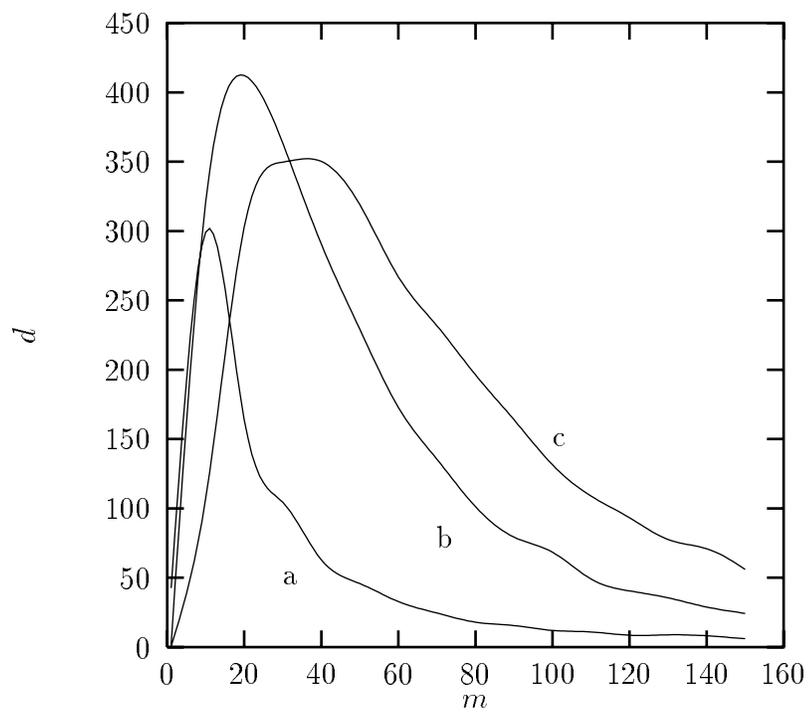
\item [Figure 4] The graph of the asymptotic difference between the number of
affected vertices in random and far edge immunization, $d$ as function
of number of vertices immunized in one time step, $m$. The three curves 
(a), (b) and (c) are for ${\cal F} = 0.0022$, $0.0084$ and
$0.0162$ respectively. The curve (b) corresponds to small-world network.
The other parameters are as in Fig. 3.
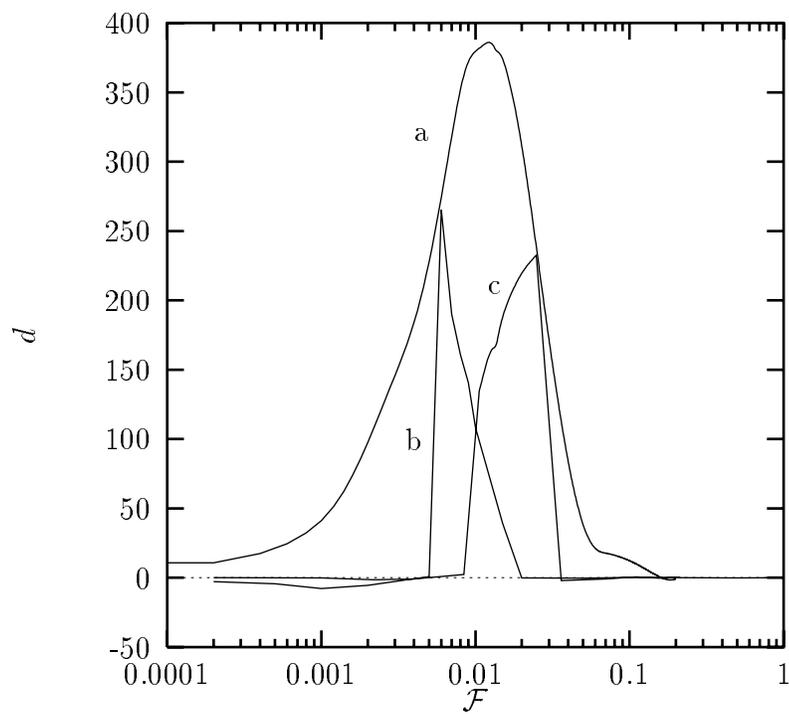
\item [Figure 5] The graph of $d$ as function of ${\cal F}$. The three 
curves (a), (b) and (c) are plotted for $m = 30$, $10$ and $80$
respectively. The figure shows that the immunization method suggested
here is most effective in small-world networks.
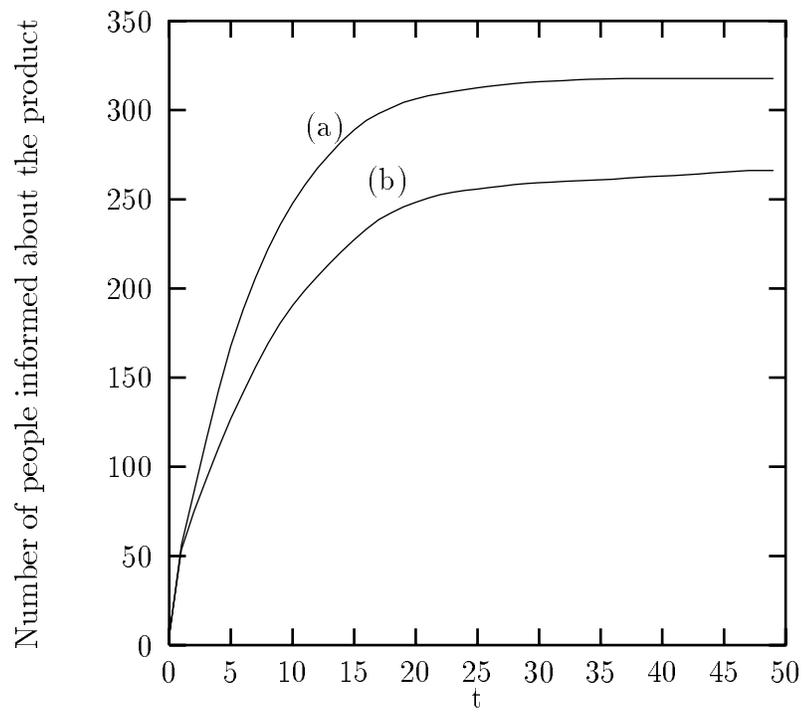
\item [Figure 6] The graph of number people informed as function of
$t$. The curves (a) and (b) show the result for far edge centers
and random centers respectively. The simulation is carried out on a
small-world network with 1000 vertices and 10000 edge. The initial
advertisement is done from five centers. The probability function $q_t$ is
chosen as $q_1 = 0.8$ and $q_i = 0.18$, where $i \geq 2$.
\end{description}
%\end{multicols}
\end{document}